\documentclass{ws-ijmpb}

\begin{document}

\markboth{Giorgi and Busch}
{Genuine correlations in finite-size spin systems}

%
\catchline{}{}{}{}{}
%

\title{GENUINE CORRELATIONS IN FINITE-SIZE SPIN SYSTEMS  }

\author{GIAN LUCA GIORGI} 

\address{Department of Physics, University College Cork\\ Cork, Republic of Ireland\\
ggiorgi@phys.ucc.ie}

\author{THOMAS BUSCH}

\address{Department of Physics, University College Cork\\
 Cork, Republic of Ireland\\
 Quantum Systems Unit, Okinawa Institute of Science and Technology\\ Okinawa 904-0411, Japan\\
thbusch@phys.ucc.ie}

\maketitle


\begin{abstract}
Genuine multipartite correlations in finite-size XY chains  are studied  as a function of the applied external magnetic field. We find that, for low temperatures, multipartite correlations are sensitive to the parity change in the Hamiltonian ground state, given that they exhibit a minimum every time that the ground state becomes degenerate. This implies that they can be used to detect the factorizing point, that is, the value of the external field such that, in the termodynamical limit, the ground state becomes the tensor product of single-spin states.  
\end{abstract}

\keywords{Spin chains; multipartite correlations; ground state factorization.}

\section{Introduction}

 Quantum spin chains have been extensively studied in
the context of  quantum statistical mechanics, with special interest directed at phase transitions  at zero temperature, which represent a purely quantum effect.\cite{sachdev} During last years, the behavior of entanglement around quantum critical points  in various systems has become a huge subject of investigation.\cite{osborne,osterloh,amicoRMP} Even if quantum phase transitions take place in the termodynamical limit, where superselection rules force the system to spontaneously break its symmetry, a finite-size analysis can be used to foresee the existence of quantum critical points. Another interesting property of quantum spin chains is the existence of a special value of the external field, located within the ordered
symmetry-broken phase, such that the ground state turns out to be factorized. While around the factorizing field there is no change of symmetry, it separates two regions where symmetry breaking takes place due to two different mechanisms.\cite{barouch}

On the other side, the interest toward quantum correlations different from entanglement  has been stimulated by the possibility of achieving  quantum speed-up using separable (unentangled) states. A remarkable example is represented by the so-called deterministic quantum computation
with one qubit (DQC1) protocol.\cite{zurek,henderson,DQC1,datta} 
Quantum discord is especially important given that it is able to capture quantum correlations in states that are not entangled.\cite{zurek,henderson} It has been shown that it is present in almost all quantum
states,\cite{acin2010} and the relation between discord and entanglement has
been discussed.\cite{cornelio,streltsov,piani}  In contrast to entanglement,
discord can be generated using local noise,\cite{steve,bruss,ciccarello1,ciccarello2} and it is not
monogamous.\cite{monogamy} Beyond the conjectured role in computational speed-up, quantum discord has other operational interpretations, since it quantifies the amount of unlockable classical correlations\cite{madhokcavalcanti}  and  accompanies the emergence of quantum synchronization.\cite{sync} 


The generalization of quantum discord to multipartite systems has followed different routes.\cite{rulli,chakrabarty} In Ref.~\refcite{genuine}, it was proposed to use the relative entropy to quantify the amount of genuine quantum and classical correlations, that is, the amount of correlations that cannot be accounted for by considering any of the possible sub-partitions of the whole system.

The analysis of quantum correlations beyond entanglement in quantum spin chains has been  discussed for instance, in Refs.~\refcite{steve,sarandy,maziero,laura,tomasello}.
 In this original paper, we consider a finite XY chain in a transverse field and focus our attention on the genuine correlations in the system.

  The paper is organized as follows. In Sec. \ref{model}, we introduce the model and discuss its solution; in Sec. \ref{gc}, we review the indicators of genuine correlations and in   Sec. \ref{results} the results are presented. Finally, Sec. \ref{concl} contains our conclusions.

\section{Model}\label{model}

The following Hamiltonian describes a chain of $N$ coupled spins (XY chain) in the presence of a  homogeneous external field:
\begin{equation}
H= -J\sum_{n=1}^{N} \left[(\frac{1+\gamma}{2})\sigma_n^x\sigma_{n+1}^x+(\frac{1-\gamma}{2})\sigma_n^y\sigma_{n+1}^y\right]- h \sum_{n=1}^{N} \sigma_n^z,
\end{equation}
where $\sigma_n^\alpha$ ($\alpha=x,y,z$) are the usual Pauli matrices, $\gamma$ quantifies the anisotropy in the $XY$ plane, $h$ is the transverse magnetic field, and  boundary conditions are
 imposed by defining $\sigma_{N+1}^\alpha\equiv \sigma_1^\alpha$.  In the following, we will assume $J=1$ and use it as an energy scale.  The analytical solution of the model can be obtained using the Jordan-Wigner transformation,\cite{katsura,lieb} which maps spins into spinless fermions.

The above Hamiltonian is invariant under the $Z_{2}$ group of the rotations by $\pi $ about the $z$ axis, given that it commutes with the parity operator $P=\prod_{l}\sigma _{l}^{z}$.
As a consequence, eigenstates of $H$ are classified depending on the
parity eigenvalue. This system undergoes a quantum phase
transition at the critical point $h_{C}=1$.\cite{barouch} Below this value, 
in the thermodynamic limit,
the odd and the even lowest eigenstates become degenerate, the Hamiltonian symmetry is spontaneously broken, and spontaneous magnetization along the $x$
axis appears.  On the other hand,
for $h>h_{C}$,  due to the existence of a non vanishing energy gap,
the ground state keeps its parity (even). 

As pointed out in Ref.~\refcite{barouch}, below the critical point,
two different symmetry breaking mechanisms take place. For  $h<h_{F}$ (where $h_{F}$ is defined through $h_{F}^{2}+\gamma ^{2}=1$) two-body correlation functions oscillate as a function of the spin distance, while for $h>h_{F}$, they decrease monotonically. For $h=h_{F}$, the ground state
factorizes,\cite{kurmann,verrucchi,amico,baroni,giampaolo,gianluca} i.e. it can be written as
\begin{equation}
\left| \Psi
_{F}^{\pm }\right\rangle =\otimes _{l}\left( \cos \frac{\alpha}{2} \left|
\uparrow _{l}\right\rangle \pm \sin \frac{\alpha}{2} \left| \downarrow
_{l}\right\rangle \right),
\end{equation}
 where $\alpha =\arccos \sqrt{ \left( 1-\gamma
\right) /\left( 1+\gamma \right) }$.

An explanation for the two different symmetry breaking mechanisms can be found by analyzing the finite-size solution of the problem. 
Looking at the  lowest odd and even
eigenvalues of $H$ in the symmetry broken region for finite $N$ as
a function of the transverse field, a series of 
level crossings for $h=h_{i}$ can be observed.\cite{hoeger,rossignoli} The number of such level crossing is equal to $N/2$ for $N$ even and to $(N/2)+1$  for $N$ odd (in this latter case, there is always degeneracy in $h=0$). At each $h_{i}$ the ground state changes its
symmetry, and, in the thermodynamic limit, this kind of structure
implies two different symmetry breaking mechanisms. For
$0<h<h_{F}$, as $N\rightarrow \infty $, the set $\left\{
h_{i}\right\} $ of the degeneracy points becomes a denumerable
infinity, while for $h_{F}<h<1$ there is  the usual symmetry
breaking due to the vanishing of the gap.  Ref.~\refcite{amico} has  pointed out that the two regions are characterized by two qualitatively different kinds of bipartite entanglement: for $h<h_F$, there is antiparallel entanglement (the main contribution comes from the antiparallel Bell states), while for $h>h_F$, there is parallel entanglement (parallel Bell states are predominant).

\section{Genuine correlations}\label{gc}

in Ref.~\refcite{modi},    Modi \textit{et al.} proposed  to measure a property of a state as the distance between the state itself and its closest state without that property. Following this principle, total, classical, and quantum correlations in a state $\rho$ can be measured by means of the relative entropy, defined as $S(\rho||\sigma)={\rm Tr}(\rho \log \rho -\rho \log \sigma)$, between $\rho $ and the closest relevant states $\sigma$.  In Ref.~\refcite{genuine},  it was suggested to use  the relative entropy  as an indicator of genuine correlations.

 The total information (or correlation information) of a $N$-partite state $\varrho$ is given by
\begin{equation}
 {\cal T}(\varrho)=\sum_{i=1}^N S(\varrho_n)-S(\varrho),\label{total}
\end{equation}
where $\varrho_n$ is the reduced density matrix of subsystem $n=1,2,\dots,N$ and $S(x)=-{\rm Tr}\{x \log x\}$ is the von Neumann entropy. ${\cal T}$ is equal to the relative entropy between $\varrho$ and its closest product state 
(which does not contain any correlation) $\pi_\varrho=\varrho_1 \otimes \varrho_2 \otimes \dots \otimes \varrho_N$:\cite{modi}
\begin{equation}
{\cal T}(\varrho)=S(\varrho\parallel\varrho_\pi)={\rm Tr} \{\varrho (\log\varrho -\log\pi_\varrho)\}.
\end{equation}
${\cal T}^{(N)}$, the genuine part of ${\cal T}$, that is, the part of correlations that contains all the contributions that cannot be accounted for by considering
any of the possible subsystems separately, is measured by the relative entropy between $\varrho$
and the closest state without $N$-partite correlations, i.e. the closest state which is factorized at least along a bipartite cut.\cite{genuine} Then,  the bipartite cut along which the closest state is found  divides the system into two partitions (let us call them $a$ and $b$). As shown in Ref.~\refcite{modi}, 
the minimum distance occurs when $\varrho_{a}$ 
and $\varrho_{b}$ are given by 
the marginals of the total state $\varrho$, that is, when  $\varrho_{a}={\rm Tr}_b \varrho$ and $\varrho_{b}={\rm Tr}_a \varrho$. Then,
\begin{equation}
{\cal T}^{(N)}(\varrho)=S(\varrho\parallel\varrho_{a}\otimes\varrho_{b})=S(\varrho_{a})+S(\varrho_{b})-S(\varrho). 
\end{equation}
The reduction of the problem of finding genuine correlations in multipartite systems to an effective bipartite problem offers a double advantage. From one side, genuine classical and quantum correlations can be defined simply extending the definitions (involving minimization procedures) given in bipartite systems;\cite{zurek,henderson} from the other side, genuine total correlations are a usual bipartite mutual information and are then very easy to compute.

\section{Results}\label{results}

\begin{figure}
\begin{center}
\includegraphics[width=10cm]{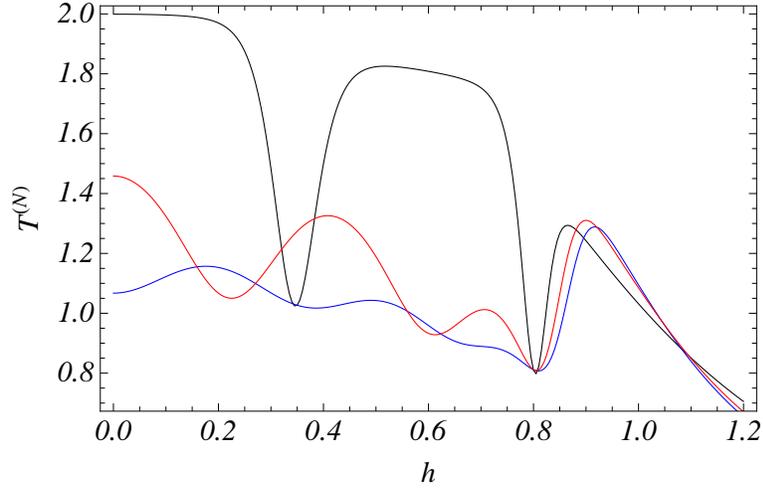}
\caption{Total genuine correlations for four spins (black line), six spins (red line),  and seven spins (blue line). The temperature is $T=0.01$ and the anisotropy parameter is $\gamma=0.6$. This value of $\gamma$ implies that the factorizing field is $h_F=0.8$. This corresponds to the last minimum of ${\cal T}^{(N)}$.}
\label{fig1} 
\end{center}
\end{figure}

We now  present the behavior of total genuine correlations for short XY chains, where finite-size effects are more evident. In Fig. \ref{fig1}, we plot  ${\cal T}^{(N)}$ as a function of the transverse field $h$ for chains of four, six, and seven sites and for very low temperatures, where the thermal state is very close to the ground state. As we can see, ${\cal T}^{(N)}$ shows four minima for seven sites, three minima for six sites, and two minima for a chain of four spins. It can actually shown that these minima fall in the level crossing $h=h_{i}$ between the lowest odd and even eigenvalues of $H$ and the last level crossing is the factorizing field. Therefore, the symmetry change of the ground state deeply influences the way in which total genuine correlations act in response of the external field. 

While the results we are presenting refer to total correlations, in the case of zero temperature, where the system is found in its ground state, genuine classical (${\cal J}^{(N)}$) and quantum (${\cal D}^{(N)}$) correlations display the same behavior, given that, in such case, $ {\cal J}^{(N)}={\cal D}^{(N)}={\cal T}^{(N)}/2$.\cite{genuine} The calculation of ${\cal J}^{(N)}$ and ${\cal D}^{(N)}$ for finite temperatures would require a numerical minimization over the set of all possible measurements.\cite{zurek,henderson} Nevertheless, for relatively small temperatures, we expect the same qualitative behavior as  ${\cal T}^{(N)}$.  A peculiar behavior of (bipartite) quantum discord around $h_F$ has been also observed in Ref.~\refcite{tomasello}. Spin-spin quantum discord turns out to be independent of the distance between the spins.  

\begin{figure}
\begin{center}
\includegraphics[width=10cm]{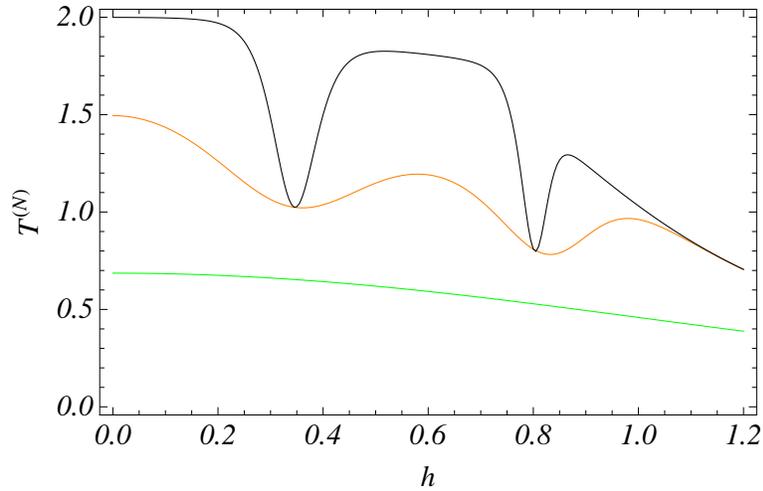}
\caption{Total genuine correlations in a chain of four spins for different temperatures. The black line corresponds to $T=0.01$, the orange line refers to $T=0.05$, while the green line corresponds to $T=1$. The anisotropy parameter is $\gamma=0.6$. }
\label{fig2} 
\end{center}
\end{figure}

On the other hand, by increasing the temperature the system state becomes highly mixed, and the effect described before is rapidly washed out. In Fig. \ref{fig2}, we plot ${\cal T}^{(N)}$ as a fuction of the transverse field for three different temperatures ($T=0.01$, $T=0.05$, and $T=1$). In the last case, there is no evidence of the dependence of ${\cal T}^{(N)}$ on the ground state level crossing.

We stress that, since we are considering finite-size systems, the symmetry is not spontaneously broken and the plots  presented here refer to symmetric states. While for the symmetry-broken factorized state the absence of correlations is obvious, the fact that, around $h_F$, correlations in symmetric states have a minimum is not an intuitive result.  

\section{Conclusions}\label{concl}

In conclusion, we have studied the behavior of genuine correlations for  finite-size XY chains in a transverse external field. We have found that, for zero temperature, genuine correlations exhibit minima for values of the external field where the odd and even ground states are degenerate. Then, the  change of symmetry sector of the ground state reflects itself in the behavior of genuine quantum and classical correlations. A special role is played by the factorizing field, which turns out to be the last minimum observed. On the other hand, at least from the finite-size analysis, at the critical point genuine correlations  do not display any distinctive feature.

\section*{Acknowledgements}
This project  was supported by Science Foundation Ireland under project number 10/IN.1/I2979. We would like to thank Steve Campbell for valuable discussions.

\end{document}